\documentstyle [12pt,graphicx,doublespace] {article} 
\makeatletter
\renewcommand\section{\@startsection {section}{1}{\z@}%
                                   {-4ex \@plus -1ex \@minus -.2ex}%
                                   {4ex \@plus.2ex}%
                                   {\normalfont}}
\renewcommand\subsection{\@startsection{subsection}{2}{\z@}%
                                     {-4ex\@plus -1ex \@minus -.2ex}%
                                     {4ex \@plus .2ex}%
                                     {\normalfont}}
\renewcommand\subsubsection{\@startsection{subsubsection}{3}{\z@}%
                                     {-4ex\@plus -1ex \@minus -.2ex}%
                                     {4ex \@plus .2ex}%
                                     {\normalfont}}
\makeatother

\newcommand {\rz} % 
{\ifmmode {I\hskip -3pt R} \else {\hbox {$I\hskip -3pt R$}}\fi}  
\begin{document}
\textwidth=14cm
\textheight=21.5cm
\pagestyle{plain}
{\noindent\small revised: July 11, 2003}   % submit date
%--------------------- Title ----------------------------------------
\begin{singlespace}
\vspace*{0.6cm}\noindent
A VEHICULAR TRAFFIC FLOW MODEL BASED ON A STOCHASTIC ACCELERATION PROCESS
\vspace{24pt}\\
K.T. Waldeer 
\vspace{24pt}\\ 
Department of Transportation and Traffic\\
Braunschweig/Wolfenb\"uttel University of Applied Sciences\\
Karl-Scharfenberg-Str. 55, 38229 Salzgitter\\
Germany\\
Th.Waldeer@Fh-Wolfenbuettel.de
%-------------------- abstract ---------------------------------------
\section*{ABSTRACT}
 A new vehicular traffic flow model based on a stochastic jump process in 
vehicle acceleration and braking is introduced. It is based on a master 
equation for the single car probability density in space, velocity and 
acceleration with an additional vehicular chaos assumption and is
derived via a Markovian ansatz for car pairs.   
This equation is analyzed using simple driver interaction models in the 
spatial homogeneous case. Velocity distributions in stochastic
equilibrium, together with the car density dependence of their moments,
i.e. mean velocity and scattering and the fundamental diagram are presented.
\end{singlespace}
%---------------------------------------------------------------------
\section{INTRODUCTION}
Since the beginning of measurement of the vehicular flow on roads, it
is well known that the dependencies of flow quantities 
show characteristic features mainly independent on individual drivers.
Therefore a wide
spread of theoretical models are
developed in the last decades. Today these models are roughly sorted
into three categories: microscopic, macroscopic and mesoscopic
models, see, e.g., \cite{Ash66,Leu88,May90,KKW96}. In microscopic modeling the 
acceleration of each car is
given as a function of a subclass of all other considered kinematic
variables, e.g., \cite{Pip53,GHR61,Wie74,BHN*94}. There, driver behavior is 
introduced using deterministic
parameters or stochastic functions. Those models
are often solved by computer simulations, e.g., 
\cite{Hel61,FL67,SSN*95,KWG97,Lud98}. The aim of macroscopic modeling is
building time propagation equations for the measured flow variables,
i.e. car density, traffic density and mean velocity at each place of the road, e.g.,
\cite{LW55,Ric56,Pay71,New93,Hel95b}. These
propagation equations are often of differential type using densities as 
continuous limits of the discrete traffic flow \cite{Dag97}. Vehicle number 
conservation on a road segment gives rise to the
kinematic equation, which propagates the car density in analogy to
the differential mass conservation law in fluid dynamics. The flow or
mean velocity is modeled either by using a special ansatz or by
constructing a second, dynamic equation, like the Navier-Stokes ansatz
in fluid dynamics. Up to now there is an open discussion of the
applicability of dynamic equations to the description of traffic
flow \cite{Dag95,DCB98,AR00}. \par
Another class of models follows the assumption that traffic flow is a
stochastic process \cite{Ada36}. Here queuing theory methods or 
state probability distribution master equations are introduced, e.g., 
\cite{Mil61,PH71,AN95,Hel97a}. Because nearly all models use
information of the  
single
vehicle behavior as input and produce results of the whole traffic
flow, this model type is often called mesoscopic.
The mesoscopic ansatz was applied to single vehicle states, e.g., 
\cite{PA60,Pav75,Phi77,Hel95b,Nel95,WK96},
as well as for vehicle clusters \cite{BK99,MP97}. Their derivation is analog 
to the equations in gas kinetic transport theory,
especially the Boltzmann transport equation, though there is no
momentum and energy conservation in the interaction between two 
vehicles. Models based on single car state
distributions are widely called kinetic models in traffic flow. \par
There are two principle binary interaction types in kinetic models. The
acceleration of a vehicle due to the lack of leading vehicles and the
braking due to a leading vehicle. Overtaking normally is introduced via
a special passing probability depending on the car density, \cite{KKW96}, or 
using multi lane coupling \cite{HG97}.  
Though acceleration often is modeled as a mean-value function 
\cite{Pav75,Hel97a},
the de acceleration interaction is a jump process in the velocity
variable, i.e. vehicles have infinite braking strength with vanishing
durance. At the end of a de acceleration interaction the following car has the
same velocity as the leading one. Kinematic variables of the
leading car remain unchanged.
Due to the velocity jump process kinetic models include the implicit
assumption that the interaction durance is much smaller than the mean
free driving time between two interactions. Additionally a 'vehicular
chaos assumption' must be fulfilled \cite{Pav75,Nel95}.  
Car following experiments on road segments show that
 velocity changes due to interactions with leading cars take 
 several seconds typically \cite{Hoe72}, where moderate
 acceleration/braking values in low or medium dense
 traffic are assumed. Under these conditions during an interaction another
 interaction can occur, producing multiple interactions, which are 
not included in those kinetic models. This effect increases with
increasing car density because the overtaking probability
decreases. Therefore kinetic models are often applied to partly constrained 
traffic \cite{Leu88}. \par
There are several possibilities to enhance kinetic models into the
higher car density regime. One of them is to introduce a continuous
velocity change due to a mean acceleration and a mean velocity
scattering. Then the single car state probability density is
described by a 
Vlasov-Fokker-Planck type equation \cite{Hel97a,IKM03}. Another
possibility is to introduce the acceleration variable into the
underlying stochastic process to be the main driver control variable
of the interaction, changed in a discontinuous manner. The
second  ansatz is followed in this paper.\par  
The traffic flow model introduced in the following sections is a
master equation approach for the single car state probability density
function similar to kinetic models. The main difference to these
models is the jump process in the acceleration variable, whereas the
velocity changes continuously and depends on the acceleration in a
deterministic way. As is shown  
in car following measurements acceleration changes take in the order
of one second \cite{Hoe72}. So except in extreme cases the assumption of 
binary interactions seems to be valid even at higher densities. Acceleration 
changes depend not only on vehicle kinematics but  also on the driver skill 
and behavior \cite{CH80}, which introduces an independent component to the 
model and therefore
justify the vehicular chaos assumption.
\par
The model is based on a car following ansatz, using kinematic variables of 
vehicle pairs, building a stochastic vector, propagating in time as a
Markov process  
following the Feller-Kolmogorov equation \cite{Fel71}. This
equation is reduced into an equation for calculating the single car state 
probability density by generalization of the vehicular chaos theorem used in 
kinetic models \cite{Nel95,WK96}. For model evaluation a simple interaction 
function between the cars of a leading car pair is constructed and
applied to  a homogeneous 
traffic flow without overtaking. The influence of different interaction 
functions to the velocity distribution and the existence of stochastic
equilibria together with the car density dependence of the typical
flow features are discussed.  \par
Note that a lot of material used in the following section is standard
in mathematics of stochastic processes, although some repeating is
important to clarify the underlying model assumptions.   
%------------------------------------------------------------------------
\section{DERIVATION OF THE MODEL EQUATION}
This section is divided into two subsections. First, the stochastic process is 
defined and second, using a vehicular chaos assumption, a master equation 
for the state probability density of single cars is constructed.
%........................................................................
\subsection{Definition of the Stochastic Process}
For a given car pair at time $t$ the leading car is located at $\bar{x}_t$ 
with velocity $\bar{v}_t$ and acceleration $\bar{a}_t$ and the following car 
is located at $x_t<\bar{x}_t$ with velocity $v_t$ and acceleration
$a_t$ defining a state vector ${\bf y}_t=(x_t,\, v_t,\,
a_t,\,\bar{x}_t,\,\bar{v}_t,\,\bar{a}_t)$. As time increases from $t$
to $t+\tau$, where $\tau$ is infinitesimal small, $x_t,\, v_t,\,
\bar{x}_t,\,\bar{v}_t$ changes continously to $x_{t+\tau},\,
v_{t+\tau},\, \bar{x}_{t+\tau},\, \bar{v}_{t+\tau}$ due to the
kinematic laws. Restricting the model to binary interactions, where
the leading car is not influenced by the following one,
i.e. $\bar{a}_{t+\tau}=\bar{a}_t$, the acceleration of the following
car $a_{t+\tau}$ changes discontinuously depending on the state vector
${\bf y}_t$. The jump hight then is given by
$\epsilon_{\tau}=a_{t+\tau}-a_t$. Because $\epsilon_\tau$ in general is
distributed stochastically with probability density
$p_j(\epsilon_\tau |{\bf y}_t)$ the state vector is a
random vector ${\bf Y}_t$ with representation ${\bf y}_t$ at time $t$,
therefore following a Markov process. The transfer probability density
$p$ from state ${\bf Y}_t={\bf y}_t$ to state ${\bf Y}_{t+\tau}$ is given by
\begin{eqnarray}
\label{1}
&&p( {\bf Y}_{t+\tau} | {\bf y}_t )=\nonumber \\
&&\delta (x_{t+\tau}-x_t-v_t\tau-{1\over 2}a_t\tau^2)\delta
(v_{t+\tau}-v_t-a_t\tau) 
\delta (\bar{x}_{t+\tau}-\bar{x}_t-\bar{v}_t\tau-{1\over
  2}\bar{a}_t\tau^2)
\cdot\nonumber \\
&&\delta (\bar{v}_{t+\tau}-\bar{v}_t-\bar{a}_t\tau)\delta
(\bar{a}_{t+\tau}-\bar{a}_t)\cdot \left(\delta
  (a_{t+\tau}-a_t)+p_j(\epsilon_\tau\neq 0|{\bf y}_t)\right) 
\end{eqnarray}       
where $\delta (z)$ is the Dirac function, \cite{Sch66}, representing the
deterministic, continuous parts of the change process. During $\tau$
there can be an acceleration jump due to density $p_j$ or no change,
represented by an $\delta$-function. Because both events are
incongruous, they appear in Eq.~\ref{1} as a sum in brackets. 
Note that $p_j$ can be written conditioned, $p_j(\epsilon_\tau\neq
0|{\bf y}_t)=\tilde{p}_j(\epsilon_\tau|\epsilon_\tau\neq 0,{\bf
  y}_t)\cdot P_j(\epsilon_\tau\neq 0|{\bf y}_t)$, where $\tilde{p}_j$ is the
density of a change of height $\epsilon_\tau$ under the condition that
a jump occurs and $P_j$ is the probability of a jump.\par
Defining the state probability density of a leading car pair $f_2({\bf
y}_t)=f_2({\bf y},t)$ at time $t$, the time propagation of this
quantity is described by the Feller-Kolmogorov equation \cite{Fel71}
\begin{equation}
\label{2}
{\partial\over\partial t} f_2({\bf y} ,t)=
\int_{{\bf z}\in\Omega} \left( q({\bf z}\rightarrow{\bf y},t)f_2({\bf z},t)
-q({\bf y}\rightarrow{\bf z},t)f_2({\bf y},t)\right) d{\bf z}
\end{equation}
with initial condition $f_2({\bf x},t=0)$. 
Here $q$ is the density of the transfer rate between states and is connected 
to the transfer probability density $p$, Eq.\ref{1}, as
\begin{equation}
\label{3}
q({\bf y}\rightarrow{\bf z},t)=\left.\frac{dp}{d\tau}\right|_{\tau=0^+}\; .
\end{equation}
Normally the components of the state vector ${\bf z}$ are defined in some 
given intervals producing restrictions to the integration area $\Omega$. Here 
$\Omega$ is expanded to the whole state vector space
by continuing the state probability density $f_2$ zero valued outside of the 
restricted area. So possible source terms are shifted into possible boundary 
conditions. The influence of the restrictions will be discussed below.\par     
Now Eqs.~\ref{1} and \ref{3} are inserted into Eq.~\ref{2} and some
algebra with $\delta$-functions and their derivatives are done
\cite{Sch66}. Together with the total transfer rate $Q$, defined by
\begin{equation}
\label{4}
Q({\bf y},t)=\int_{{\bf z}\in\Omega}q({\bf y}\rightarrow{\bf
  z},t) d{\bf z}=\left.{dP_j\over
  d\tau}\right|_{\tau=0^+}\; ,
\end{equation}
and the calculated integral
\begin{eqnarray}
\label{5}
 &&\int_{{\bf z}\in \Omega}q({\bf z}\rightarrow{\bf
  y},t)f_2({\bf z} ,t) d{\bf z}=
-v{\partial f_2\over\partial x}-a{\partial f_2\over\partial v}
-\bar{v}{\partial f_2\over\partial \bar{x}}-\bar{a}{\partial
  f_2\over\partial \bar{v}}\nonumber \\
&&+\int_{a^\prime\in \rz} Q(x,v,a^\prime
,\bar{x},\bar{v},\bar{a},t)\cdot \tilde{p}_j( \epsilon_0=a-a^\prime|\epsilon_0\neq
0, x,v,a^\prime ,\bar{x},\bar{v},\bar{a},t) \cdot \nonumber \\
&& \hspace{5.2cm} f_2(x, v, a^\prime ,\bar{x},\bar{v},\bar{a},t)d a^\prime\; ,
\end{eqnarray}
where ${\bf z} = 
(x^\prime ,v^\prime ,a^\prime  ,\bar{x}^\prime , \bar{v}^\prime 
,\bar{a}^\prime )$ and $\epsilon_0=\epsilon_{\tau_0=0^+}$ is used, 
the Feller-Kolmogorov equation reads as
\begin{eqnarray}
\label{6}
 &&{\partial f_2\over\partial t}
+v{\partial f_2\over\partial x}+a{\partial f_2\over\partial v} 
+\bar{v}{\partial f_2\over\partial \bar{x}}+\bar{a}{\partial
  f_2\over\partial \bar{v}}+Q f_2=\nonumber \\ 
&& =\int_{a^\prime\in \rz} Q(x,v,a^\prime
,\bar{x},\bar{v},\bar{a},t)\cdot \tilde{p}_j( \epsilon_0=a-a^\prime
|\epsilon_0\neq 
0, x,v,a^\prime ,\bar{x},\bar{v},\bar{a},t)\cdot\nonumber \\
&&\hspace{5.2cm} f_2(x, v, a^\prime,\bar{x},\bar{v},\bar{a},t)d a^\prime\; .
\end{eqnarray}
The function arguments on the left side are suppressed for simplification.
There are two important features not included in Eq.~\ref{6} until
yet. First, the state of the leading car $(\bar{x},\, \bar{v},\,
\bar{a})$ only changes continously. So it has not the same character
than a following car, where acceleration jumps can occur. Second, the
equation does not include  the property
$x<\bar{x}$ for a leading car pair. Both features will be introduced
in the next section via a vehicular chaos 
theorem, splitting $f_2$ partially in a product of single car
probability densities.
%.............................................................................
\subsection{Vehicular Chaos Assumption and the Master Equation}
Vehicular chaos assumptions are often used in Boltzmann type mesoscopic 
traffic flow models. The main idea is to decouple $f_2$ into a product of two 
single car probability densities neglecting correlations. In this
paper the ideas of  
R.~Wegener and A.~Klar \cite{WK96} and P.~Nelson \cite{Nel95} are essentially 
used.\par
The car pair state density $f_2$ can be written as a product of
conditional densities as
\begin{equation}
\label{7}
f_2(x, v, a,\bar{x},\bar{v},\bar{a},t)=l_1(x,v,a,t)\cdot
l_2(\bar{x}|x,v,a,t)\cdot l_3(\bar{v},\bar{a}|\bar{x},x,v,a,t)\; ,
\end{equation}
where the first factor $l_1$ is identified with the single car
probability density $f(x,v,a,t)$ and the second factor $l_2$ is the
conditioned distance density $D$ introducing the distance variable  
$h=\bar{x}-x$ instead of $\bar{x}$, i.e.
\begin{equation}
\label{8}
l_2(\bar{x}|x,v,a,t)d\bar{x}=D(h|x,v,a,t)dh\; .
\end{equation}
The third factor $l_3$ describes the behavior of the leading car depending on 
the state of the following one and therefore should be a function of $f$. The 
vehicular chaos ansatz assumes that the velocity $\bar{v}$ and the
acceleration $\bar{a}$ of the leading car is independent on the state
of its following one resulting in a special form of $l_3$,
\begin{equation}
\label{9}
l_3(\bar{v},\bar{a}|\bar{x},x,v,a,t)=l_3(\bar{v},\bar{a}|\bar{x},t)={f(\bar{x},\bar{v},\bar{a},t)\over {\cal F}(\bar{x},t)}\; ,
\end{equation}
with definition
\begin{equation}
\label{10}
{\cal F}(\bar{x},t)=
\int_{\bar{v},\bar{a}} f(\bar{x},\bar{v},\bar{a},t)\, d\bar{v}\, d\bar{a}\; .
\end{equation}
This special form of $f_2$ ensures both features not included in
Eq.~\ref{6}. Each leading car now is distributed as
$f(\bar{x},\bar{v},\bar{a},t)$ and therefore is also a following one in
another leading car pair. The distance density $D$ is zero valued for
distances $h<h_{\min}$, which is equal to the mean car length. It is a
behavioral function and therefore does not depend explicit on a
special place 
$x$ at time $t$. However, there is an additional dependence of the
distance on the overall flow, mainly represented by its mean values
like mean velocity, scattering of velocity and acceleration,
etc.. Typically these are lower moments of the single car state probability $f$
and so space and time dependent. They are included into $D$ as
elements of a
moment vector ${\bf m}_f(x,t)$ of $f$, i.e. $D=D(h|v,a,{\bf
  m}_f(x,t))$.
Note that this moment vector can also include further flow dependent
functions like the mean distance, which are not direct moments of $f$ (
see Eq.~\ref{12a} below).
 \par
Inserting this vehicular chaos assumption in Eq.~\ref{6}, integrating over 
$\bar{a}$, $\bar{v}$ and $\bar{x}$ or $h$ and bearing in mind that the single 
car state density $f$ vanishes at the infinite boundaries of $\bar{v}$ and 
$\bar{x}$ continuing $f$ like $f_2$, the resulting time propagation equation 
for $f$ is given by
\begin{eqnarray}
\label{11}
&& {\partial f\over\partial t}
+v{\partial f\over\partial x}+a{\partial f\over\partial v}=
\int_{h,\bar{v},\bar{a},a^\prime}dh\, 
d\bar{v}\, d\bar{a}\, da^\prime\cdot (1-P_o)\cdot{
  f(x+h,\bar{v},\bar{a},t)\over{\cal F}(x+h,t)}\cdot\nonumber \\
&&\left\lbrace 
\sigma(a|x,h,v,\bar{v},a^\prime,\bar{a},t)Q(x,h,v,\bar{v},a^\prime,\bar{a},t)
D(h|v,a^\prime,{\bf m}_f(x,t)) f(x,v,a^\prime,t)\right.\nonumber \\
&& 
-\left.\sigma(a^\prime|x,h,v,\bar{v},a,\bar{a},t)Q(x,h,v,\bar{v},a,\bar{a},t) 
D(h|v,a,{\bf m}_f(x,t)) f(x,v,a,t)\right\rbrace\nonumber \\
\end{eqnarray}
introducing
\begin{equation}
\label{12}
 \tilde{p}_j(\epsilon_0=a-a^\prime= |\epsilon_0\neq 0, x,v,a^\prime 
,\bar{x},\bar{v},\bar{a},t)=(1-P_o) 
\sigma(a|x,v,a^\prime,h,\bar{v},\bar{a},t)\; .
\end{equation}
Here a simple ansatz for overtaking, often assumed in kinetic traffic models 
\cite{KKW96} is used. It is based on the overtaking probability
density $P_o$, which depends on the kinematic variables
of the car pair and other vehicles nearby. Because of its mathematical 
complexity, in literature this function often is simplified on the local 
car density only. For a more precise modeling of overtaking
effects a  
multilane model has to be constructed in the same manner as is done
elsewhere \cite{HG97}.\par
The car density $K$ at place $x$ and time $t$ is given by the inverse of
the mean distance $\bar{H}$ at the same place and time,
i.e. $K(x,t)=1/\bar{H}(x,t)$. This unconditioned mean value can be
calculated using the vehicular chaos ansatz resulting in  
\begin{eqnarray}
\label{12a}
\bar{H}(x,t)&=&\int_{v,a,\bar{x},\bar{v},\bar{a}}|\bar{x}-x|{f_2(x, v,
a,\bar{x},\bar{v},\bar{a},t)\over {\cal F}(x,t)} \,d\bar{x}\, d\bar{v}\, d\bar{a}\, dv\,
da=\nonumber \\
&&\int_{v,a,h>h_{\min}} h\cdot D(h|v,a,{\bf m}_f(x,t))
{f(x,v,a,t)\over{\cal F}(x,t)} dh\, dv\, da \; .
\end{eqnarray} 
 Together with Eq.~\ref{12a}, Eq.~\ref{11} describes the whole
traffic dynamic. Note that if $D$ is independent on the state of the
car $(v,a)$, Eq.~\ref{12a} is always fulfilled. \par
Eq.~\ref{11} has the same structure as those, used in Enskog theory of dense 
gases \cite{Ens17} or other kinetic traffic flow models in literature 
\cite{KKW96,Hel97a} with a spatial correlation via $D$. It couples microscopic 
car pair interactions defined by $\sigma$, $Q$ and $D$ with macroscopic 
traffic flow quantities, which are moments of the state probability density 
$f$. The main differences to other traffic flow models based on kinetic type 
equations are the additional, independent acceleration variable, free
functions  
describing the car pair interaction and the model derivation from a
Markov process. In contrast to the often used mean acceleration ansatz,
here the acceleration is used as the control variable of the driver
process as in reality \cite{Pav75}.\par
Note that using the same derivation technique a model based on a continuous 
stochastic change of acceleration can be derived, which is similar to the 
Fokker-Planck-equation framework \cite{Gar97}, but in the acceleration
variable. There, in contrast to the jump 
process the time sequence of acceleration changes due to interactions has to 
be known as input, making the model more complex and less applicable 
\cite{Wal00a}.\par  
For solving Eq.~\ref{11} an additional initial condition and spatial
and velocity boundary conditions  must 
be specified. The velocities 
of all cars have to be non negative, i.e. $v\geq 0$, and the acceleration $a$ 
is bounded $a_{\min}\leq a\leq a_{\max}$, where $a_{\min}<0$ is the maximal 
braking value. If a speed limit is defined on a given road segment,
$w(x)$, then all velocities are lower than this limit $v\leq
w(x)$. Normally the spatial coordinate $x$ lies between a minimal and
a maximal value, where in flow and out flow conditions must be specified.
These restrictions also influences the construction of
the interaction functions prohibiting values outside of the allowed
intervals. Concrete interaction examples for model testing will be
given below. \par 
Driver measurements show that interaction decisions are mainly done
by the own driving state $v,a$ and the distance $h$ and roughly the relative
velocity $\bar{v}-v$ between the cars in a leading car pair
\cite{Mic63,ER73,May90}; the acceleration of the leading car is not
taken into account.\par
Assuming a spatial homogeneous road segment and the arguments
discussed above, Eq.~\ref{11} is reduced
to a Boltzmann type equation
\begin{eqnarray}
\label{13}
&& {\partial f\over\partial t}+a{\partial f\over\partial v}=
\int_{\bar{v},a^\prime}d\bar{v}\, da^\prime\cdot
\tilde{f}(\bar{v},t)\cdot\nonumber \\
&&\quad\left\lbrace 
\Sigma(a|v,\bar{v},a^\prime ,{\bf m}_f(t))f(v,a^\prime,t)
-\Sigma(a^\prime|v,\bar{v},a,{\bf m}_f(t))f(v,a,t)\right\rbrace\, , 
\end{eqnarray}
with $\tilde{f}(v,t)=\int_{a} f(v,a,t)\,da$ and 
\begin{equation}
\label{14}
\Sigma(a|v,\bar{v},a^\prime ,{\bf m}_f(t))=\int_{h_{\min}}^\infty 
\sigma(a|h,v,\bar{v},a^\prime)Q(h,v,\bar{v},a^\prime)
D(h|v,a^\prime,{\bf m}_f(t))\, dh \; 
\end{equation}
is the weighted interaction density.
Before the complex spatial dynamics of a traffic flow is studied in a
model, the typical phase diagrams in stochastic equilibrium are
calculated and proofed to be in agreement with measured data. In
contrast to gas kinetic, where the universal Maxwell-Boltzmann
distribution exits, in traffic flow it depends on chosen interaction
functions, because there are no {\em Sto\ss invarianten} in the
process. The simplest way to calculate solutions in stochastic
equilibrium is starting from an homogeneous initial condition and
following its time propagation assuming a homogeneous road segment
using Eq.~\ref{13}.  Naturally the real process can not assumed to
stay homogeneous during time propagation, because fluctuations in
spatial dynamics occur, vanishing when reaching the equilibrium.  
%----------------------------------------------------------------------------
\section{SOME REMARKS ON THE MACROSCOPIC MODEL BEHAVIOR}
In contrast to stochastic equilibrium, the spatial dynamics in traffic
flow models are often
studied using macroscopic flow equations calculated from the base
equation using the moment method. 
Although this paper deals with direct solutions in the equilibrium,
some preliminary remarks on the macroscopic model behavior are given
to facilitate a comparison to other models. Additionally for a better
understanding of the model structure and comparison to other kinetic
models the equation for the reduced state probability density
$\tilde{f}(x,v,t)=\int_a f(x,v,a,t)\, da$ is discussed. This equation
is calculated integrating Eq.~\ref{11} with reference to $a$. At first
sight, the result
\begin{equation}
\label{15}
{\partial\tilde{f}\over\partial t}+v{\partial\tilde{f}\over\partial x}+
{\partial A\tilde{f}\over\partial v}=0
\end{equation}
seems to be a Vlasov-like equation with a mean acceleration field
$A$. This function is calculated by the first moment of
Eq.~\ref{11},
\begin{equation}
\label{16}
{\partial A\tilde{f}\over\partial t}+v{\partial A\tilde{f}\over\partial x}+
{\partial S\tilde{f}\over\partial v}=\int_a a\cdot {\cal J}[f]\, da\, ,
\end{equation}
 where $S=\int_a a^2f(x,v,a,t)\, da$ and the operator $\cal J$ is the
 right side of Eq.~\ref{11}. So $A$ depends not only on the kinematic
 state of the car, but also  on the higher moments of $f$
 especially on the scattering $S$. Now $S$ can be calculated by the
 next moment of $f$, etc.. For practical use a closure condition must
 be introduced resulting in a system of differential equations. As
 mentioned above in literature the Vlasov-Fokker-Planck ansatz is often
 used to model acceleration behavior in a mean value sense with 
 velocity scattering. In contrast to those models, here an
 explicit velocity diffusion term can not be generated straight
 forward, but it is implicit included into the function $A$ via
 Eq.~\ref{16}. Up to now it is an open question, under which conditions
 an explicit diffusion term can be constructed in a consistent way.\par
 Macroscopic equations are produced by velocity {\em and} acceleration
 moments only depending on $x$ and $t$. Integrating Eq.~\ref{15} with
 reference to $v$, using Eq.~\ref{10} yields the continuity equation
\begin{equation}
\label{17}
{\partial{\cal F}\over\partial t}+{\partial V{\cal F}\over\partial x}=0.
\end{equation}
The equation for the mean velocity $V(x,t)$ results analogous
calculating the first velocity moment   
\begin{equation}
\label{18}
{\partial V\over\partial t}+V{\partial V\over\partial x}=
B-{\partial \Theta\over\partial x}-{\Theta\over{\cal F}}\cdot{\partial {\cal
      F}\over\partial x},
\end{equation}
with mean acceleration $B(x,t)=\int_v A(v,x,t)dv$ and velocity
variance 
$\Theta(x,t)=\int_v v^2\tilde{f}(v,x,t)dv- V^2(x,t)$.
$B$ and $\Theta$ can be calculated by the next moments of Eqs.~\ref{15}
and~\ref{16}. Mean acceleration
$B$ and mean velocity $V$ are coupled by the mean covariance and the
acceleration scattering in the second velocity moment of
Eq.~\ref{16}, etc.. The structure of both first order
Eqs.~\ref{17},~\ref{18} 
are similar to  
those known from miscellaneous literature models \cite{Zha99}. There $B$ is
specified by an ansatz in contrast to this model. Up to now closure
conditions needed for practical use of the macroscopic equations are
not developed and therefore have to be done in a future work. 
%----------------------------------------------------------------------------
\section{EXPLICIT SOLUTIONS IN STOCHASTIC EQUILIBRIUM}
In this section the model equation is solved for special interaction
functions. The aim is to
show that under justifiable input the result also seems to be
applicable. Additionally the section shows the existence of equilibria
for these exemplary interactions. First, a simple relative velocity
dependent interaction, which can be found in low density traffic flow
is developed and its solution behavior is discussed. Second, a simple
distance threshold interaction is introduced and the car density
dependence of the flow quantities is shown.
\subsection{Relative Velocity Dependent Interactions in a Low Density
  Traffic Flow}
At low car densities, the occurrence of an interaction together with
the interaction strength are not depending on the distance between the
cars, but on the change of the distance, i.e. the relative velocity
mainly \cite{ER73}. The driver of the following car has nearly no
quantitative estimates on the kinematic state of the leading car and
therefore decides his new acceleration value by his own
behavior. Only the sign of the new acceleration value, braking or
accelerating, depends on the sign of the relative velocity. The
simplest approximation of this behavior is a two value acceleration
change, given by
\begin{equation}
\label{19}
 \sigma(a|\bar{v}-v)=\Theta(\bar{v}-v)\delta(a-a_1)+\Theta(v-\bar{v})\delta(a
-a_2)\; ,
\end{equation}
where $\Theta(z)$ is the Heaviside step function and $a_1 > 0$ is the
fixed new acceleration value and $a_2 < 0$ 
is the fixed de acceleration value.
Assuming this relative velocity dependent interaction, the interaction
rate $Q$ is only a function of $\bar{v}-v$. In this model all driver
behave similar and therefore have nearly the same velocities. So $Q$ can
be Taylor expanded for small relative velocity values up to first order as
\begin{equation}
\label{20}
Q(\bar{v}-v)=Q_0+r_0|\bar{v}-v|\; .
\end{equation}
In the following, two extreme cases will be considered: the relative
velocity rate $Q_0=0$, which in gas kinetic is called hard sphere
rate, and the constant rate $r_0=0$, which in gas kinetic is called a
Maxwell rate. Because in this section a homogeneous traffic flow based
on Eq.~\ref{13} is supposed, Eq.~\ref{14} is given by 
\begin{equation}
\label{21}
\Sigma (a|\bar{v}-v)=Q(\bar{v}-v)\cdot\sigma(a|\bar{v}-v)\; ,
\end{equation}
which is independent on the chosen concrete distance density
$D$. Therefore the velocity distribution is also independent on
distance features. To produce car density dependencies, the solution
$f(v,a,t)$ must be inserted into Eq.~\ref{12a} and a concrete $D$ must
be defined there.\par
Note that this interaction profile does not ensure positive velocities. So it
can only be applied, if the mean velocity of the flow is large against
its velocity variance, which lets the probability for self-defeating negative
velocities vanish approximately. In the next subsection it is shown,
how such negative velocities can be avoided using velocity boundary
conditions.\par
Assuming an initial condition, where all cars have one of the two
possible acceleration values, the state probability density can be
 written as
\begin{equation}
\label{22}
f(v,a,t)=f_1(v,t)\delta(a-a_1)+f_2(v,t)\delta(a-a_2)\; .
\end{equation}
$f_1$ and $f_2$ are the state densities of the accelerating and the braking 
cars resp.. Inserting Eq.~\ref{21} and Eq.~\ref{22} into Eq.~\ref{13} results 
in two equations
\begin{eqnarray}
\label{23}
{\partial f_1\over\partial t}+a_1 {\partial f_1\over \partial v}&=&
f_2(v,t)\int_{\bar{v}>v}Q(\bar{v}-v)\tilde{f}(\bar{v},t)
d\bar{v}\nonumber \\
&-&f_1(v,t)\int_{\bar{v}<v}Q(\bar{v}-v)\tilde{f}(\bar{v},t)
d\bar{v}\; ,\nonumber \\
{\partial f_2\over\partial t}+a_2 {\partial f_2\over \partial v}&=&
f_1(v,t)\int_{\bar{v}<v}Q(\bar{v}-v)\tilde{f}(\bar{v},t) d\bar{v}\nonumber\\
&-&f_2(v,t)\int_{\bar{v}>v}Q(\bar{v}-v)\tilde{f}(\bar{v},t) d\bar{v}\; .
\end{eqnarray}
From both equations it can be shown that the solutions after a
sufficient long period of time $t$ are given asymptotically by
\begin{equation}
\label{24}
f_i(v,t)\sim {1\over 2}\tilde{f}_{\mathrm eq}\left(v-A\cdot t\right)\, , \quad i=1,2\, ,
\end{equation}
where $A=(a_1+a_2)/2$ is the mean acceleration and  $\tilde{f}_{\mathrm eq}$ is the equilibrium solution of the same system,
Eq.~\ref{23}, whereas 
$a_1$ is replaced by $(a_1-a_2)/2$ and $a_2$ is replaced by
$(a_2-a_1)/2$, . Note that for these new acceleration values the mean
value $A$ vanishes and so a stochastic equilibrium exists.\par
   Starting from an initial condition for $f_i$, after some time the shape of
the functions are changed into the shape of $\tilde{f}_{\mathrm eq}$,
which then propagates along the velocity axis with mean
acceleration $(a_1+a_2)/2$.  Figure~\ref{f1} shows an example of
this time evolution starting with two Gaussian shape densities and
resulting in a moving one given by Eq.~\ref{29} below. It is
calculated numerically by discretizing Eq.~\ref{23} using a standard explicit
discretizing scheme for the left hand side and the standard trapezoidal
rule for the integrals on the right hand side \cite{Wal00a}.\par 
To calculate the equilibrium solution, first $a_1=-a_2=a_0$ is set.
For $\partial f_i/\partial t=0$ and using $\tilde{f}_{\mathrm
  eq}(v)=f_1(v)+f_2(v)$, which is calculated directly from Eq.~\ref{22}, the
sum of both Eqs.~\ref{23} results in an equation for the
equilibrium velocity density   
\begin{equation}
\label{25}
{d\tilde{f}_{\mathrm eq}(v)\over dv}={\tilde{f}_{\mathrm eq}(v)\over a_0}\left(
\int_v^{\infty} Q(\bar{v}-v)\tilde{f}_{\mathrm eq}(\bar{v}) d\bar{v}-
\int_{-\infty}^v Q(\bar{v}-v)\tilde{f}_{\mathrm eq}(\bar{v}) 
d\bar{v}\right)\;  
\end{equation}
with additional condition $\int_v \tilde{f}_e(v) dv =1$. 
The difference of both equations results in $f_1(v)=f_2(v)$ and
so $f_i(v)=\tilde{f}_{\mathrm eq}(v)/2$ for $i=1,2$.
The general solution, Eq.~\ref{22}, is
\begin{equation}
\label{25a}
f(v,a)=\tilde{f}_{\mathrm eq}(v){\delta(a-a_0)+\delta(a+a_0)\over 2}\; .
\end{equation}
For rate $Q=r_0|\bar{v}-v|$ Eq.~\ref{25} is reduced to
\begin{equation}
\label{26}
{d\tilde{f}_{\mathrm eq}(v)\over dv}={\tilde{f}_{\mathrm eq}(v)\over
  a_0}(V-v) 
\end{equation}
with a Gaussian probability density as solution
\begin{equation}
\label{27}
\tilde{f}_{\mathrm eq}(v)={1\over
  \sqrt{2\pi}}\,{1\over\sigma_v}\,e^{-(v-V)^2/2\sigma_v^2}\; .
\end{equation}
Here the mean velocity $V$ is a free parameter and
$\sigma_v=\sqrt{a_0/r_0}$. 
In the case of a constant rate $Q=Q_0=1/T$, Eq.~\ref{25} is reduced to
the system
\begin{equation}
\label{28}
{d\tilde{f}_{\mathrm eq}(v)\over dv}={\tilde{f}_{\mathrm eq}(v)\over
  a_0T}\, s(v)\; , \quad {ds(v)\over dv}=-2\tilde{f}_{\mathrm eq}(v)\; .
\end{equation}
Taking the derivative of the second equation, inserting the first and
then the second again, results in $s^{\prime\prime}(v)=[s(v)^2]^\prime/2Ta_0$,
which can be solved using the method of variable separation. After
inserting $s(v)$ in the second Eq.~\ref{28}, the
solution is
\begin{equation}
\label{29}
\tilde{f}_{\mathrm eq}(v)={\pi\over 4\sqrt{3}\cosh^2\left({\pi\over2\sqrt{3}}{v-V\over\sigma_v}\right)}
\end{equation}
with free parameter $V$ again and
$\sigma_v=a_0T\pi/\sqrt{3}$. Figure~\ref{f2} shows both velocity
distributions in standardized form. The deviations are small and the
Maxwell type density is slightly less peaked. 
Both distributions agree
qualitatively to measured distributions and other theoretical work
\cite{Gre35,Hel97a}. Note that the equilibrium velocity distribution
seems not to be strong influenced by the functional form of $Q$ in this
traffic flow regime. This result is the same as is found for thermalization 
processes governed by the homogeneous Boltzmann equation in kinetic gas theory 
\cite{WU91}. There also the shape of the energy distribution during an
equilibration process is nearly independent on the interaction rate; the 
interaction rate influences mainly the timescale of thermalization.
%----------------------------------------------------------------------
\subsection{A Simple Distance Threshold Interaction Model}
\label{ss2}
In this subsection a distance oriented interaction model is
introduced, to study the car density behavior of the equilibrium
distribution. Here a simple distance interaction threshold
$H(v)=\alpha\cdot v+h_{\min}$ well established in literature
\cite{KW99b} is defined. Only if the distance 
between the cars in a leading car pair is equal to $H(v)$ an
acceleration change is done by the following car. The driver of the
following car checks this condition with a constant rate $Q=Q_0=1/T$. 
Like in the last subsection, here again a two value acceleration change
profile is used for simplicity, i.e.
\begin{equation}
\label{30}
 \sigma(a|h,v)=\Theta(h-H(v))\delta(a-a_0)+\Theta(H(v)-h)\delta(a
+a_0)\; .
\end{equation}
If the distance $h$ is larger than $H(v)$, then the following car
accelerates with $a=a_0$, otherwise it de accelerates with
$a=-a_0$. Because $\sigma$ depends on $h$, a concrete distance
correlation $D$ is needed. As discussed in literature the exponential
ansatz, unconditioned on $(v,a)$, seems to be the right choice for
moderate car densities 
\cite{Nel95}   
\begin{equation}
\label{31}
 D(h)={1\over\bar{H}-h_{\min}}\,e^{-{h-h_{\min}\over \bar{H}-h_{\min}}}\,\Theta(h-h_{\min})\; ,
\end{equation}
where $\bar{H}$ is the mean distance between the cars in a leading car
pair and therefore is equal to $/K$. Note that this ansatz fulfills
Eq.~\ref{12a}.   
Inserting this interaction profile into Eqs.~\ref{13},~\ref{14} analog
to the last section, esp. using Eq.~\ref{25a}, the equation for the
equilibrium velocity distribution is 
\begin{equation}
\label{32}
{d\tilde{f}_{\mathrm eq}(v)\over dv}={\tilde{f}_{\mathrm eq}(v)\over
  a_0T}(1-2P(h\leq H(v)))\; ,\mbox{ with} \; P(h\leq
  H(v))=\int_{h_{\min}}^{H(v)} D(h)\, dh\; . 
\end{equation}
The interaction profile does not avoid the unrealistic negative
velocities. For $v<0$ there is $H(v)<h_{\min}$ and therefore
$P(h\leq H(v))=0$. The solution of Eq,~\ref{32} then is an exponential
function decreasing with decreasing $v$. For $v>0$ Eq.~\ref{32}
together with Eq.~\ref{31} can be solved analytically. Both solutions
$v<0$ and $v>0$
are matched at $v=0$, because there the function is continuous. After some
ponderous but straight forward calculations including the unity norm,
the probability density is given by
\begin{equation}
\label{33}
\tilde{f}_{\mathrm eq}(v)={e^{-{|v|\over a_0T}-2\beta e^{-{v\over
  a_0T\beta}\Theta(v)}}\over a_0T\left(e^{-2\beta}+\beta (2\beta
  )^{-\beta} \gamma(\beta ,2\beta)\right)}\; ,
\end{equation}
where $\gamma(x,y)$ is the incomplete gamma function and
$\beta=(\bar{H}-h_{\min})/(\alpha a_0T)$ the scaling parameter of the
solution. It compares the mean distance between two cars with the
distance change during $T$. From this solution analytic expressions for the
mean velocity and the velocity scattering can be obtained. They are
not included here, because of their mathematical complexity. The
intrinsic model error $P(v<0)=\int_{-\infty}^0 \tilde{f}_{\mathrm
  eq}(v)\, dv$ vanishes exponentially proportional to $(e/2)^{-\beta}/\sqrt{\beta}$ for large
$\beta$ and so for small car densities.
%-----------------------------------------------------------------------.
\subsection{The Introduction of Velocity Boundary Conditions}
\label{ss3}
To eliminate negative velocity probabilities absorbing boundary
conditions at $v=0$ in analogy to the Bose gas state condensation can
be constructed. If a braking car reaches $v=0$, it changes
independent on the state of the leading car his acceleration value to
$a=0$. Introducing a maximum free flow velocity $w$, the same ansatz
is possible there. If a car reaches $v=w$, its acceleration value is
set to zero. Using both boundary conditions the new state
density $\hat{f}(v,a,t)$ is calculated from
\begin{eqnarray}
\label{34}
\hat{f}(v,a,t)&=&f(v,a,t)\chi(0<v<w)\nonumber \\
&+&\int_{-\infty}^0\tilde{f}(v,t)\,dv\,\delta(v)\delta(a)+
\int_w^{\infty}\tilde{f}(v,t)\,dv\,\delta(v-w)\delta(a)\; .
\end{eqnarray}
$\chi(0<v<w)$ is the index function, which equals one for $0<v<w$ and
zero elsewhere. This boundary condition influences the
behavior of velocity distributions especially in the free flow regime
and near $v=0$. It also changes the equilibrium behavior
together with the existence statements presented in the last
subsection.  So Eq.~\ref{23}, its solution~\ref{24} together with the boundary condition
results now in an equilibrium solution $\hat{f}_{\mathrm
  eq}(v,a)=\delta(a)\delta(v)$ for $A<0$ and $\hat{f}_{\mathrm
  eq}(v,a)=\delta(a)\delta(v-w)$ for $A>0$. \par
Applying the boundary condition to the distance threshold model,
subsection~\ref{ss2},
$\tilde{f}_{\mathrm eq}(v,a)$ is given by Eq.~\ref{25a}
together with Eq.~\ref{33}. Integrating Eq.~\ref{34} with reference to
$a$ results 
the new velocity distribution $\hat{f}_{\mathrm eq}(v)$, which is
shown in figure~\ref{f3} for different car density values $K=1/\bar{H}$
and two different acceleration values $a_0$. The distributions show
the typical shape as mentioned above and their situation depends
strongly on the car density. Note that $a_0$, which is also equal to
the acceleration scattering (acceleration noise, ACN) in this simple
interaction model mainly influences the width or scattering of the
distributions. \par
The car density dependence of the mean velocity $V=\int_0^w
v\hat{f}_{\mathrm eq}(v)dv$ and the 
velocity scattering calculated from $\sigma_v^2=\int_0^w
v^2\hat{f}_{\mathrm eq}(v)dv-V^2$ are shown in the figures~\ref{f4}
and \ref{f5}. At low car densities the mean velocity equals the free
flow velocity $w$ due to the boundary condition. With increasing car
density the mean velocity shows the typical decrease.
Because for $\alpha=1.8$s an interaction
occurs at larger threshold distances than for $\alpha=1.2$s, the
decrease occurs at lower car densities. Velocity scattering results are
much more seldom in literature than mean value results. The model
shows a steep increase at moderate densities, where the transition 
from free flow to interaction oriented flow occurs. Then with
increasing car density the velocity scattering diminishes. Other
threshold oriented kinetic models show the similar behavior
\cite{KW99b}. The peak in the scattering shifts to higher car density
values when $\alpha$ decreases, which corresponds to the begin of the
mean velocity decrease in fig.~\ref{f4}. The fundamental diagram,
figure~\ref{f6} shows its typical behavior. The interaction oriented
part of the curve depends on the $\alpha$ value. Let the lower
$\alpha$-value is identified to a more aggressive driver and the
higher $\alpha$-value to a more conservative one, a traffic flow of a
mixture of both produce traffic densities in the area between both
curves. This could be a first guess to multivalued traffic flows in
the framework of this model as discussed in literature
\cite{Hog99,IKM03}.\par
In gas kinetic models interactions produce velocity jumps based on
infinite acceleration and de acceleration change values. In the model
discussed in the last two subsections this corresponds to the case of
$a_0\rightarrow\infty$. Applying this limit operation to the boundary
condition for all three solutions Eq.~\ref{27}, \ref{29} and
\ref{33}, together with Eq.~\ref{25a}, a limit probability  density
independent on the concrete interaction profile but based on velocity
jumps can be calculated
\begin{equation}
\label{35}
\hat{f}_{\mathrm eq}(v,a)=\left({1\over 2}\delta(v)+{1\over 2}\delta(v-w)\right)\cdot\delta(a)\; .  
\end{equation}
The velocity part of this density is well known in kinetic traffic
flow theory and used there as an argument for the existence of bimodal velocity
distributions \cite{Nel95,IKM03}. There, \cite{Nel95}, the factor
$1/2$ is replaced by
a more complex function of the distance probability at $v=w$, but this
difference is mainly due differences in the interaction model.
The bimodal limit distribution in principal can be derived from a more
universal qualitative argument. As long as the velocity scattering
depends on the 
acceleration scattering in a monotone increasing way, the limit case
together with the boundary conditions always produce this kind of
solution with perhaps different prefactors.
Its practical applicability seems to
be problematic, because typical peak acceleration scattering values are in
the order of 1m/s$^2$, far away from this limit case \cite{Win79}. 
 %---------------------------------------------------------------------
\section{CONCLUSIONS}
\begin{itemize}
\item[(1)]
In this paper a vehicular traffic flow model based on a stochastic car pair 
process with acceleration jumps is introduced. The model equation for the 
single car distribution function in the spatial, velocity and acceleration 
variable is of Enskog type using a vehicular chaos assumption. The model input 
is based on car following information, especially for the interaction rate, 
the interaction strength and the distance distribution. 
The behavior of solutions for different interaction models and
macroscopic moment equations are discussed. Velocity distributions in
stochastic equilibrium are calculated for special interaction cases.   
In contrast to Boltzmann theory, the existence  and shape of an
equilibrium solution of  
the model depends on the construction of the interaction functions.  
\item[(2)]
The spatial homogeneous equation for a discrete two
value acceleration change, depending only on the relative velocity,
is discussed. There an equilibrium solution exists exclusive, if the
mean acceleration vanishes. For this case velocity distributions for
constant rate and relative velocity dependent rate are calculated
analytically, showing
qualitative the same behavior as is found in measurements or other traffic 
flow models. With increasing acceleration scattering the velocity
distribution broadens, i.e. the velocity scattering increases also.   
\item[(3)]
A simple distance threshold oriented interaction model with constant
interaction rate is discussed in stochastic equilibrium. To prohibit
unrealistic negative velocities and to introduce a maximum free flow
velocity, absorbing boundary conditions in analogy to the Bose gas
state condensation are introduced. The car density dependence of the
velocity distributions, the mean velocity and velocity scattering are
shown for special interaction values. The fundamental diagram agrees
qualitatively in shape with other models.
\item[(4)]
In a next step the interaction profile must be enhanced to be more
realistic. Especially the acceleration noise, here given by a constant
value, together with the car density dependence of the
acceleration distribution must be considered. Under these conditions
the model can be solved only by computer simulation techniques. So a
stable method to solve the master equation must be developed. Further
work on this field is under way.
\end{itemize}     
%-------------------- Bibliography -----------------------------------
\begin{singlespace}
%\bibliographystyle{elsart-num}
%\bibliography{traffic}

\end{singlespace}
%------------------------ Figure Captions ----------------------------------
\newpage
\begin{singlespace}
\section*{FIGURE CAPTIONS}
\newcounter{fig}
\begin{list}{Fig. \arabic{fig}:}{\usecounter{fig}}
\item\label{f1} 
Time evolution of the partial distribution functions $f_1$ (broken lines) and $f_2$ 
(solid lines) using Eqs.~\ref{23} for Maxwell interaction with parameters 
$T=2$s, $a_1=0.4$m/s$^2$ and $a_2=-0.2$m/s$^2$. The initial conditions
at $t=0$  
are normal distributions with mean $V=28$m/s and variance
0.1m$^2/$s$^2$.
\item\label{f2}
Standardized velocity distributions in stochastic equilibrium based on 
Eq.~\ref{25}. The brocken line describes the Maxwell type interaction rate, 
Eq.~\ref{29}, where the solid line is the relative velocity type
interaction rate, Eq.~\ref{27}. 
\item\label{f3}
Equilibrium velocity distributions for different car density values
$K$ described in subsection~\ref{ss3} with
$a_0=0.05$m/s$^2$ (solid line) and $a_0=0.2$m/s$^2$ (broken line). The
following fixed
parameter values are used: $T=2.5$s, $\alpha=$1.8s, $h_{\min}=6.5$m.
\item\label{f4}
The car density $K$ dependence of the mean velocity $V$ for two
different threshold slopes as described in subsection~\ref{ss3}. The following
fixed parameter values are used: $T=2.5$s, $a_0=$0.2m/s$^2$, $h_{\min}=6.5$m.
\item\label{f5}
The car density $K$ dependence of the velocity scattering $\sigma_v$ for two
different threshold slopes as described in subsection~\ref{ss3}. The
following fixed
parameter values are used: $T=2.5$s, $a_0=$0.2m/s$^2$, $h_{\min}=6.5$m.
\item\label{f6}
The car density $K$ dependence of the traffic density $K\cdot V$ for two
different threshold slopes as described in subsection~\ref{ss3}. The
following fixed
parameter values are used: $T=2.5$s, $a_0=$0.2m/s$^2$, $h_{\min}=6.5$m.
\end{list}
\end{singlespace}
%-----------------------------------------------------------------------
\begin{center}
\begin{figure}[t]
%\centering
\includegraphics[angle=90,height=20cm]{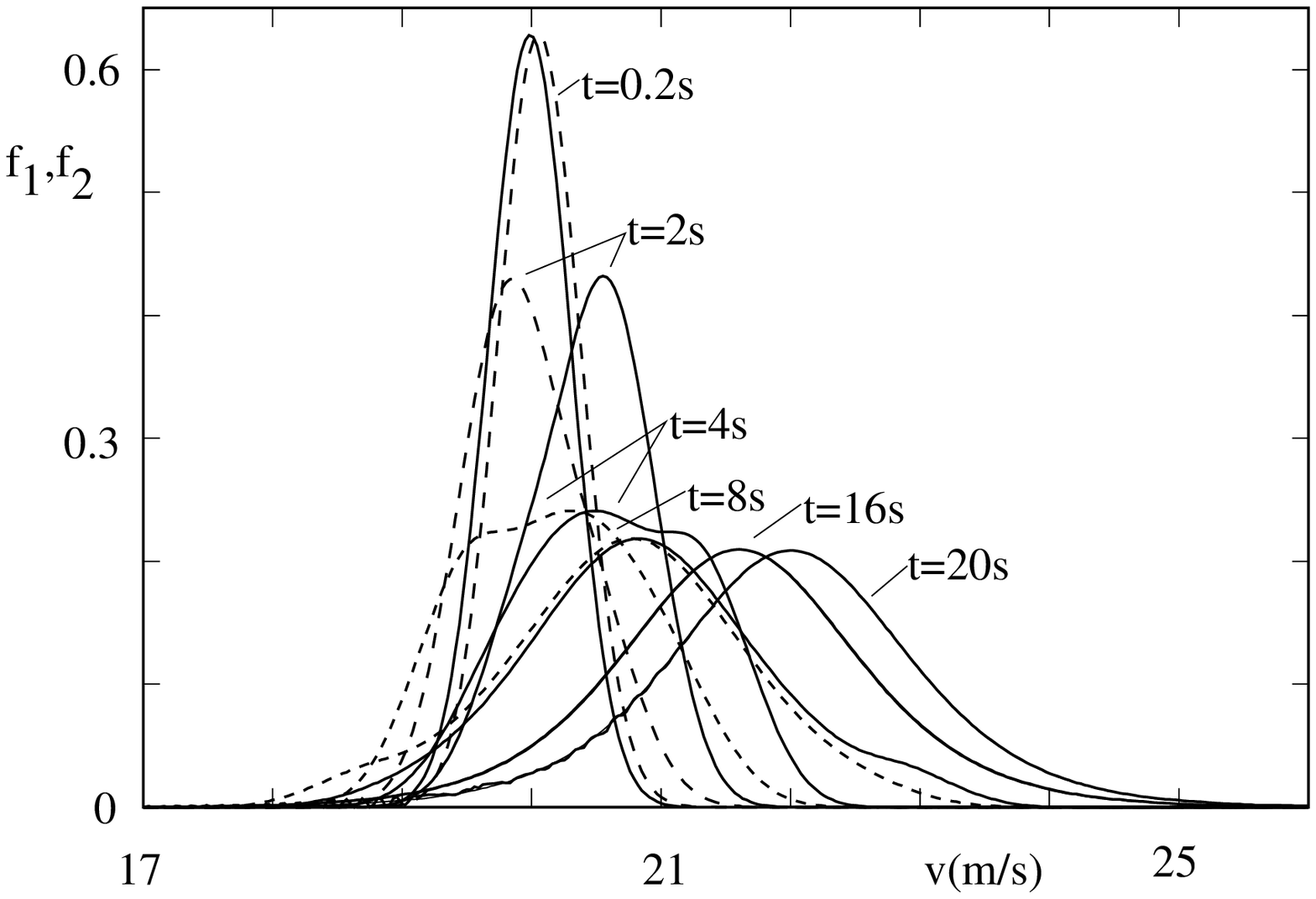}
\caption{}
\end{figure}
\newpage
\begin{figure}[t]
%\centering
\includegraphics[angle=90,height=20cm]{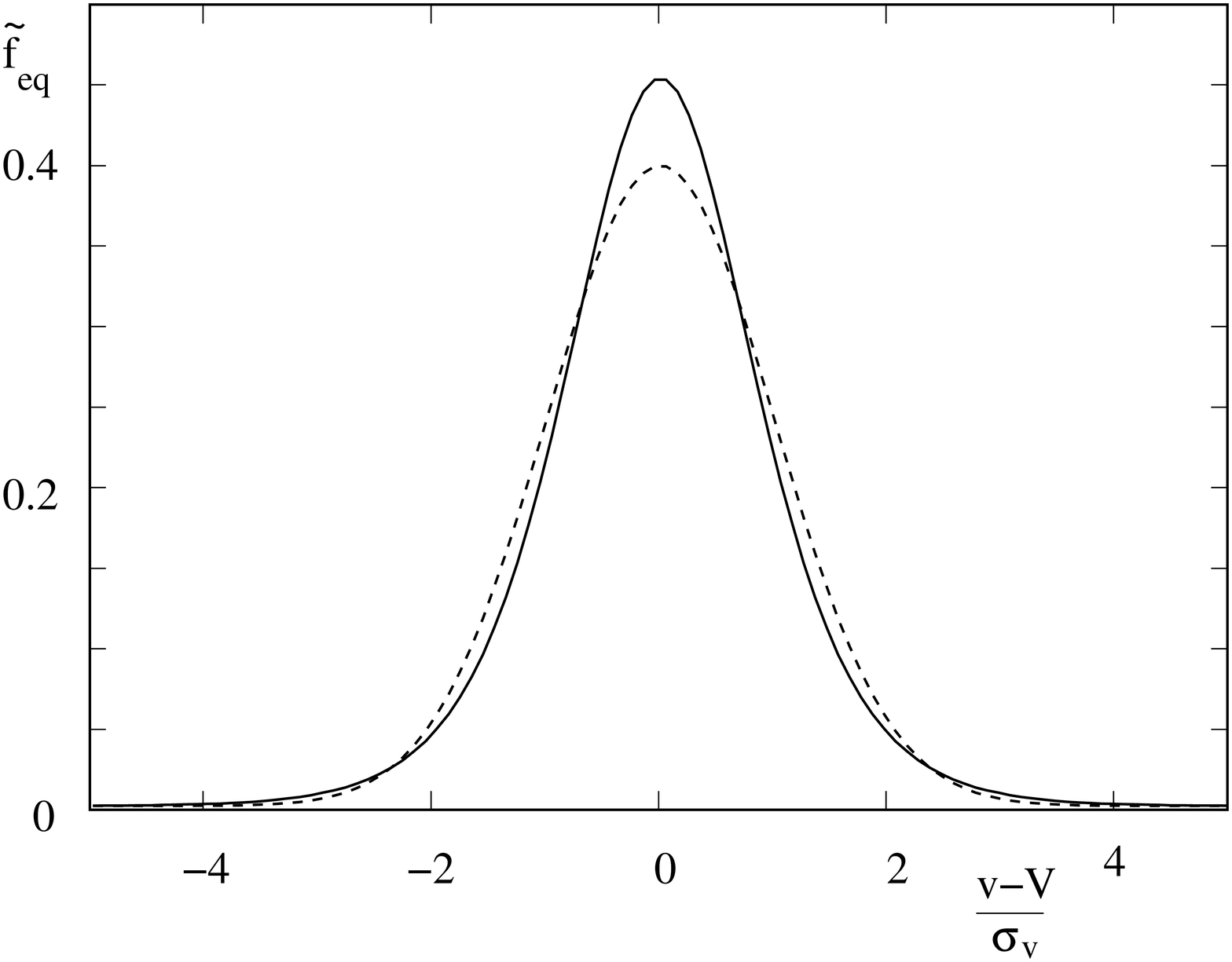}
\caption{}
\end{figure}
\newpage
\begin{figure}[t]
%\centering
\includegraphics[angle=90,height=20cm]{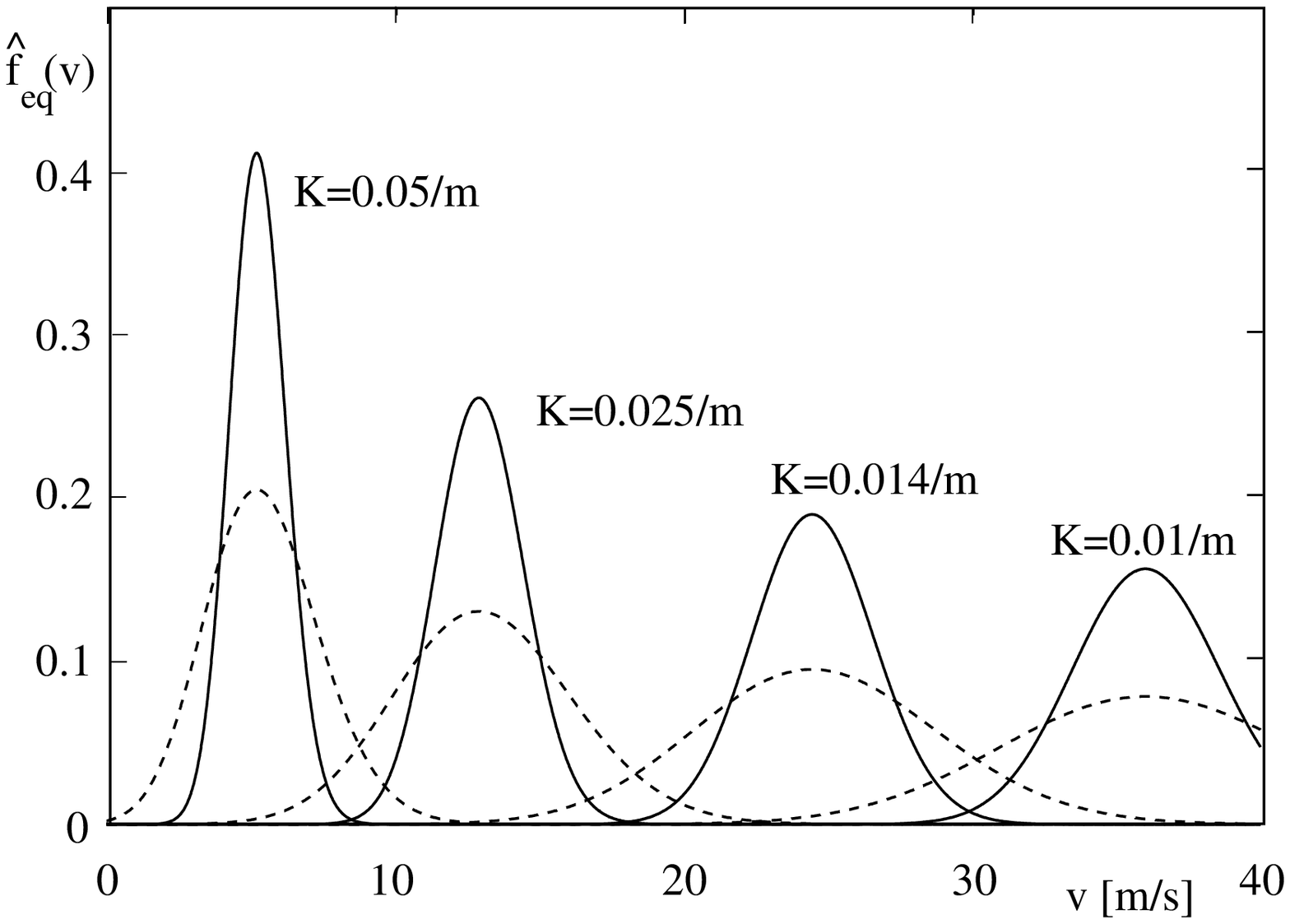}
\caption{}
\end{figure}
\newpage
\begin{figure}[t]
%\centering
\includegraphics[angle=90,height=21cm]{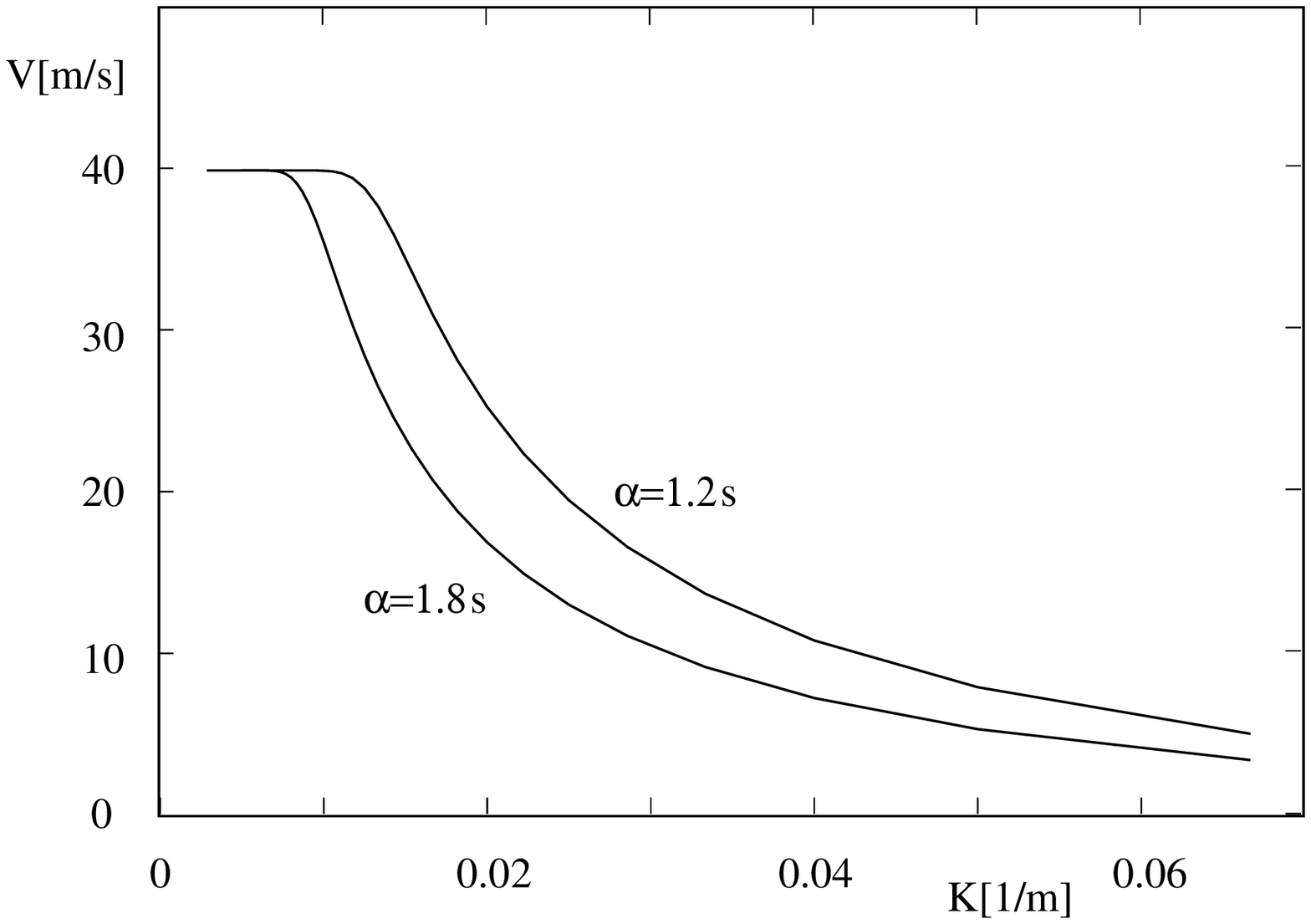}
\caption{}
\end{figure}
\newpage
\begin{figure}[t]
%\centering
\includegraphics[angle=90,height=20cm]{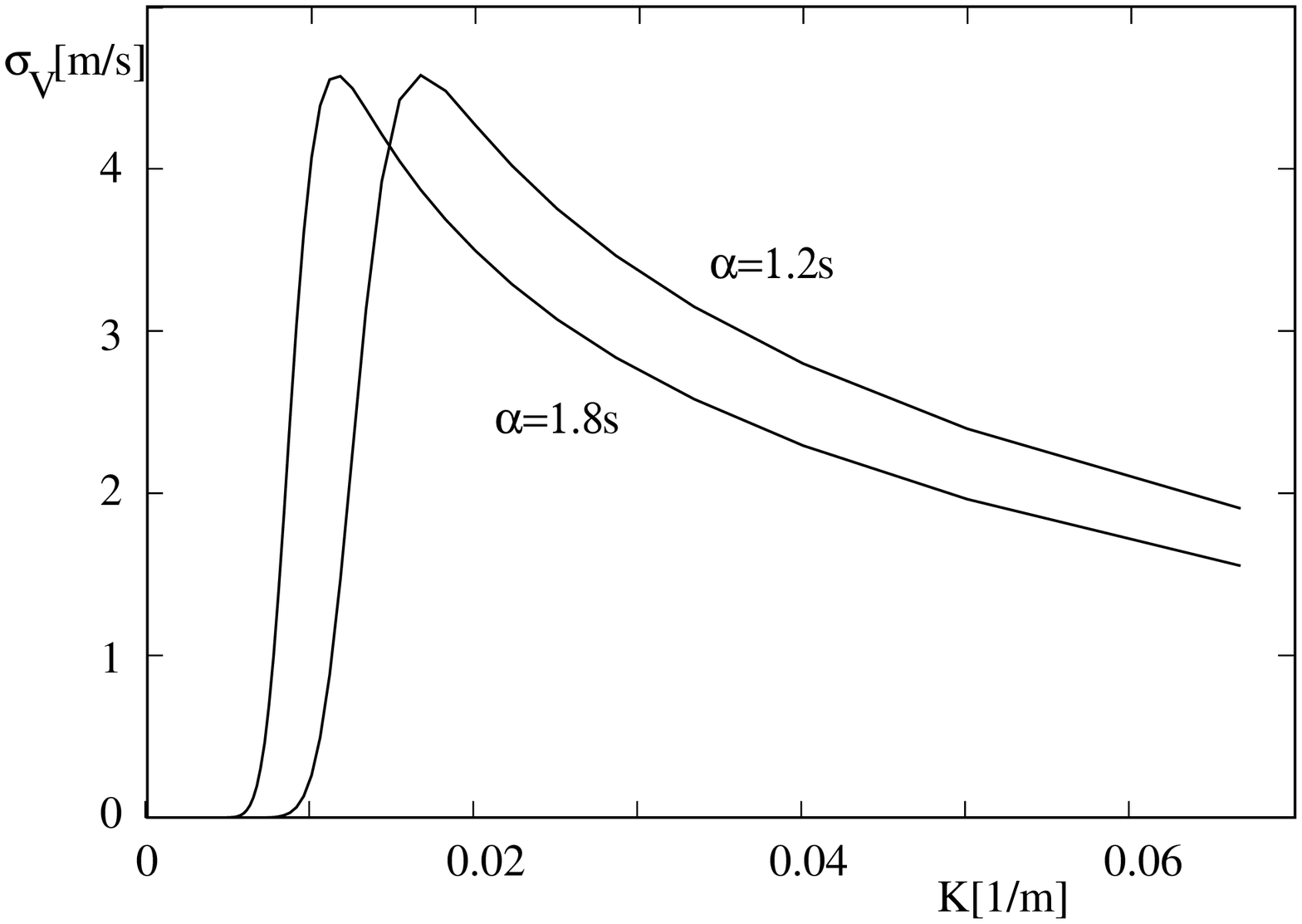}
\caption{}
\end{figure}
\newpage
\begin{figure}[t]
%\centering
\includegraphics[angle=90,height=20cm]{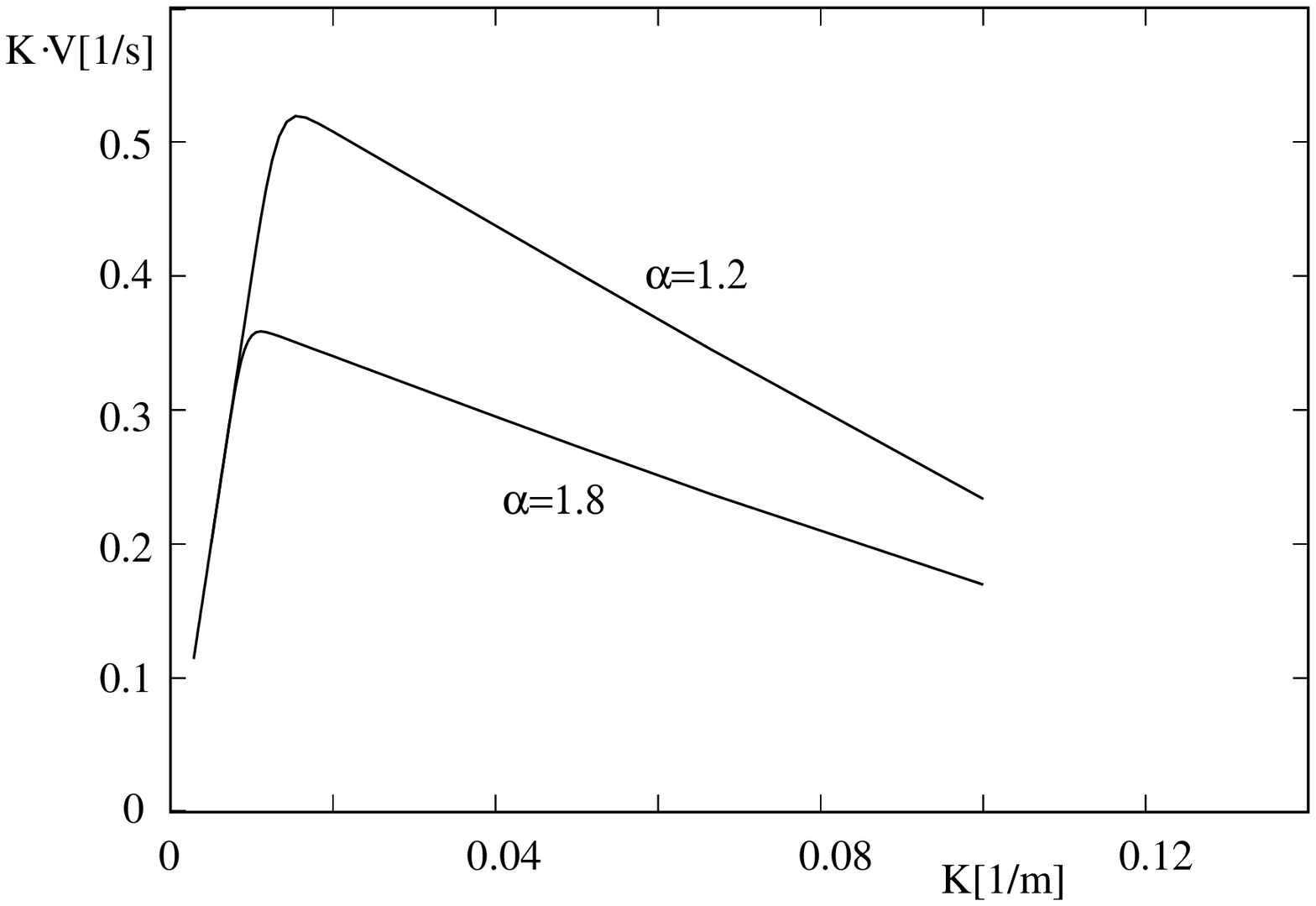}
\caption{}
\end{figure}
\end{center}
%---------------------------------------------------------------------
\end {document}